\documentclass[runningheads]{llncs}
\usepackage[T1]{fontenc}

\usepackage[utf8]{inputenc} 
\usepackage[linesnumbered,ruled,vlined]{algorithm2e} 
\usepackage{subcaption}
\usepackage{import}
\usepackage{multirow}
\usepackage{graphicx}
\usepackage{xcolor}
\usepackage{hyperref}
\usepackage{url}
\usepackage{amsmath}
\usepackage{fancyhdr}
\usepackage{booktabs} 

\usepackage{macros}

\begin{document}
\title{CoProver: A Recommender System for Proof Construction}
\titlerunning{CoProver}

\pagestyle{fancy}
\fancyhead{} 
\renewcommand{\headrulewidth}{0pt}
\renewcommand{\footrulewidth}{.5pt}
\fancyhead[LE]{\thepage~\quad~Yeh et al.}
\fancyhead[RO]{CoProver~\quad~\thepage}
\fancyfoot[C]{\footnotesize Distribution Statement A – Approved for Public Release, Distribution Unlimited}

\author{
Eric Yeh\thanks{This research was, in part, developed with funding from the Defense Advanced Research Projects Agency (DARPA). The views and conclusions contained in this document are those of the authors and should not be interpreted as representing the official policies, either expressed or implied, of the U.S. Government or DARPA.}\and
Briland Hitaj\and
Sam Owre \and
Maena Quemener \and
Natarajan Shankar 
}
\institute{
    SRI International, Menlo Park, CA 94025, USA
    \email{\{eric.yeh,briland.hitaj,sam.owre,maena.quemener,natarajan.shankar\}@sri.com}\\
}
\authorrunning{Yeh et al.}


\maketitle

\begin{abstract}
Interactive Theorem Provers (ITPs) are an indispensable tool in the arsenal of formal method experts as a platform for construction and (formal) verification of proofs. The complexity of the proofs in conjunction with the level of expertise typically required for the process to succeed can often hinder the adoption of ITPs. A recent strain of work has investigated methods to incorporate machine learning models trained on ITP user activity traces as a viable path towards full automation.  
While a valuable line of investigation, many problems still require human supervision to be completed fully, thus applying learning methods to assist the user with useful recommendations can prove more fruitful.

Following the vein of user assistance, we introduce CoProver, a  proof recommender system based on transformers, capable of learning from past actions during proof construction, all while exploring knowledge stored in the ITP concerning previous proofs. CoProver employs a neurally learnt sequence-based encoding of sequents, capturing long distance relationships between terms and hidden cues therein. We couple CoProver with the Prototype Verification System (PVS) and evaluate its performance on two key areas, namely: (1) Next Proof Action Recommendation, and (2) Relevant Lemma Retrieval given a library of theories. We evaluate CoProver on a series of well-established metrics originating from the recommender system and information retrieval communities, respectively.  We show that CoProver successfully outperforms prior state of the art applied to recommendation in the domain. 
We conclude by discussing future directions viable for CoProver (and similar approaches) such as argument prediction, proof summarization, and more.

\end{abstract}

\newcommand{\term}[1]{\MakeLowercase{\textsc{#1}}}

\setlength{\parskip}{\baselineskip}%
\setlength{\parindent}{0pt}%

\normalsize
\section{Introduction}
\label{sec:introduction}

Interactive theorem proving (ITP) is a well-entrenched technology for formalizing proofs in mathematics, computing, and several other domains.  While ITP tools provide powerful automation and customization, the task of manually guiding the theorem prover toward QED is still an onerous one.  
For inexperienced users, this challenge translates to crafting mathematically elegant formalizations, identifying suitable proof commands, and diagnosing the root cause of failed proof attempts. Whereas for the expert users, the challenge consists in navigating a large body of formalized content to ferret out the useful definitions and the right lemmas.  Both novice and expert users can benefit from recommendations in the form of proof commands and lemma retrieval that can guide proof construction.  

The goal of the present project is to scale up proof technology by
introducing CoProver as a proof recommender system that discerns suitable
cues from the libraries, the proof context, and the proof goal to offer recommendations for ITP users.
Building on the proof
technology and proof corpora of the Prototype Verification System (PVS) a state-of-the-art proof assistant \cite{owre1992pvs}, we focus on recommendations for two key tasks in ITP: Suggesting PVS commands and lemmas.  The first is to recommend the likely command an expert user would take given the current proof state.  As the number of possible commands can number over $100$, recommending steps an expert may take would be beneficial, particularly for novices.  The second is to identify lemmas for inclusion from a library of lemmas that may help with forward progress on a proof.  Currently, only lemmas from user-imported theories are considered and selection of a lemma relies on user familiarity with candidate theories and their lemmas.  For the problems PVS is commonly employed on, there are usually several hundred theories with thousands of possible lemmas combined to consider.  At this scale, even expert users with decades of experience may not be aware of all possible lemmas (or even their names) that may be relevant for their proof.  A mechanism that can automatically identify relevant lemmas at scale would be desirable.

In order to develop these two capabilities, we leverage the expert proof traces for NASA's PVS Library \footnote{\url{https://shemesh.larc.nasa.gov/fm/pvs/PVS-library/}} (PVSLib), a large collection of highly polished formal developments centered on safety-critical applications, and the PVS Prelude, a collection of theories built into PVS.  We aim to capture the expertise and intuition of the developers by training systems to emulate user decisions on these completed proofs.

Key to our approach is the use of recent machine learning techniques that can capture sequential information across a greater window than previously possible.  Indeed, we show how application of these methods over a simple sequence-based encoding of formulas can better capture relationships between proof states with commands and libraries than prior sequence encoding techniques.

We start with an overview in Section~\ref{CoProver_overview} of the CoProver system, describing the core recommendation tasks.  We describe each approach and contrast it against prior literature.  Section~\ref{sec:data_generation} contains implementation details of how sequents and states are featurized into a common representation used to provide inputs for the command prediction and lemma retrieval capabilities.  Sections~\ref{sec:command_prediction} and~\ref{sec:lemma_retrieval} provide specific details of how these are implemented, along with experiments detailing their effectiveness described in Section~\ref{sec:experiments}.  Section~\ref{sec:deployments} describes deployments of this capability on PVS, initial user evaluations of these capabilities, and  Section~\ref{sec:conclusions} concludes with a discussion of future directions.

\section{CoProver Overview}
\label{CoProver_overview}

\begin{figure}
    \centering
    \includegraphics[width=0.95\textwidth]{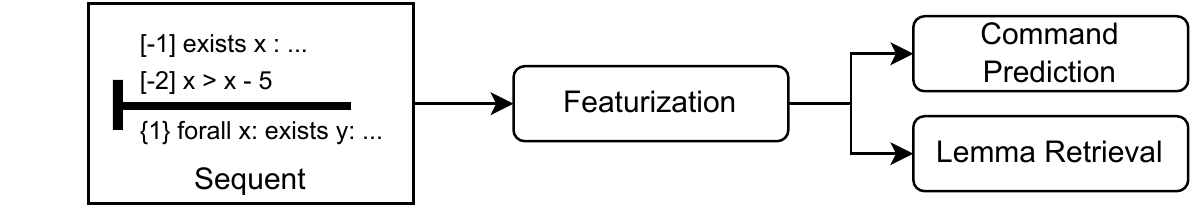}
    \caption{The CoProver system.  A sequent is featurized, and this is representation is used to provide command recommendations and retrieve relevant lemmas.}
    \label{fig:CoProver_overview}
\end{figure}

Figure \ref{fig:CoProver_overview} illustrates the CoProver system.  
The Featurization step converts the structured logical formula of the sequent and other information constituting the proof state, such as previous commands, into a token sequence that can be later ingested by the transformer model.
This sequence is provided to the Command Prediction and Lemma Retrieval modules.  Command Prediction identifies the next likely command an expert user would take in successful proofs, given similar states.  When a lemma is to be imported, Lemma Retrieval examines the state and suggests the most relevant ones from a given library, based off of a history of human-selected lemmas that have progressed their proofs.  

Both Command Prediction and Lemma Retrieval use RoBERTa\cite{liu19roberta}  to encode the proof state sequence tokens.  RoBERTa is a Transformers-based neural language model \cite{vaswani2017attention}, which can learn long-range sequences of terms useful for performing a downstream task.  Unlike $n$-grams or other Markov window methods, Transformers employ a self-attention mechanism that lets tokens employ a wider range of neighboring tokens to generate a representation suitable for a downstream task.  The token-level representations are real-valued vectors that can be tailored to perform for a variety of tasks, and have been used to give state-of-the-art performance across multiple tasks such as large language modeling, text classification, and visual understanding.
In our case, the RoBERTa-based encoding of the proof state is used as input to a multinomial classifier for predicting the next command an expert user would take.  Backpropagated error from the classifier is used to adjust, or to \emph{fine-tune} these representations make them more suitable for the classification task.

Lemma retrieval is the task of identifying which lemmas from an existing library of theories may be relevant to include to help with performing forward progress on a proof.  This task is made challenging by the fact that a potentially useful lemma may exist in a theory that the user is not aware of or might not recall the name of, and therefore would not import into the proof.  This use case has strong connections to the document-retrieval problem in information retrieval (IR) \cite{manning2008introduction}, where the goal is to identify which documents in a collection are most relevant to a search query.  A particular challenge is the fact that IR models rely on overlap and term rarity heuristics for relevance.  While the assumptions that motivate these heuristics work for natural documents and queries, whether they hold for theorem proving is in question.  To account for this, CoProver uses a training set of lemma import decisions by users to fine-tune the RoBERTa-derived proof state representations to learn combinations of proof state and lemma symbols that prove useful for identifying relevant lemmas from a given database of lemmas.

\section{Data Generation}
\label{sec:data_generation}

\subsection{Datasets}
For both command prediction and lemma retrieval, we used proof sequences from the PVSLib library, a large set of formal developments containing theorems proofs for a variety of mathematical and engineering areas.  In total PVSLib contains $184,335$ proof steps.  We note that these are completed and polished proofs so that backtracked sequences of steps are pared and only the successful sequence of proof commands and imported lemmas are retained.

\subsection{Featurization}
In order to convert the logical formulas into a representation amenable to machine learning, we convert the structured representation into a token sequence.  
While the original PVS is a well-structured formal language, featurization is still required for several reasons.  First, our models used RoBERTa to encode PVS proof states and lemmas.  As Transformer models have fixed width input, a more parsimonious encoding that strips away boilerplate while retaining the original semantics will allow longer formulas to work without truncating them.  Many formulas use names for variables and constants that are specific to that proof or theory, and having the model learn these names would reduce the generalization ability of the model.

Given this, all symbol names for functions and operators are copied over as-is.  Constant and variable names are replaced by a generic placeholder.  Integer values are retained.  Syntactic constructs such as parentheses are excluded and not included in the featurized form, as the ordering of the above can roughly capture the syntactic arrangement of the original form.  

\begin{figure}
    \centering
    \includegraphics[width=\textwidth]{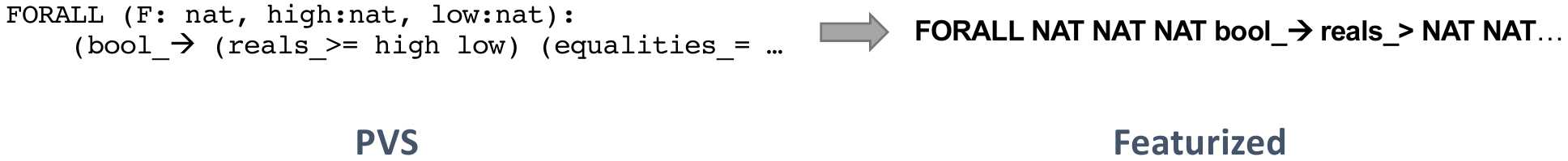}
    \caption{Featurization converts formulas in PVS (left) into a more machine learning friendly token stream (right).}
    \label{fig:featurization}
\end{figure}

An example of the input and output of this process is given in Figure \ref{fig:featurization}.  Here, symbols such as the \term{FORALL} quantifier and implication operator are preserved, while the variables \term{F}, \term{high}, and \term{low} are replaced with their type, \term{nat} representing the natural numbers.

Following previous work in Transformer-based encodings, we use special tokens \term{<ANT>}, \term{<CONS>}, and \term{<HID>} to mark regions of tokens corresponding to formulas in the sequent's antecedent and consequent.  The last corresponds to hidden formulas, a feature of PVS that allows the user to reserve certain formulas from being operated on by commands.  

For command prediction, we prefix all token streams with a \term{CMD1} special token.  This acts as a bias term for this task and can be used to condition the Transformer model for multi-task learning.  The lefthand side of Figure \ref{fig:cmdpred_seq2seq} gives an example of a featurized sequent with no antecedents and one consequent.
 
Our current setup makes the Markov assumption, that only the current state is sufficient for making our predictions.  At least anecdotally knowing which commands were performed can inform what steps are taken next, so incorporating previous commands can capture some non-Markovian information.  This is done by adding the previous three commands issued by the user to the state representation after the \term{CMD1} token.

For the lemma retrieval experiments, we modified the above procedure to allow constant and variables to be replaced with their type name.  This was done to allow matching by type, as arguments for imported lemmas also need to match by type.  Higher-order and custom types are currently represented by a placeholders.  Accounting for matches on higher-order types and on advanced type operations such as predicate-based approaches is reserved for future work.  

We use the Byte Pair Encoding (BPE) \cite{sennrich16subword,gage94bpe} from the Huggingface Transformers Library \cite{huggingface} to train a token vocabulary customized for the featurized formulas (sequents and lemmas).  BPE encodes words as a sequences of byte pairs instead of singular tokens.  This reduces the size of the vocabulary required, as rare words can be encoded using byte pairs and do not need to be memorized, while common words can be encoded in their entirety to improve efficiency.  This also introduces a tolerance for new terms, which can be assembled from suitable byte pair sequences.  
 
 As with many other Transformers based encoders, RoBERTa is first trained in a self-supervised fashion.  Given unlabeled data, one can generate labeled versions through operations where the ground truth can be determined.  Masked language modeling is one such task, where a supervised prediction problem can be constructed by masking the identity of the word, and then asking the model to predict that word \cite{devlin-etal-2019-bert}.  By conducting this type of self-supervised training on a large corpus, the resulting representations can capture distributional information about the domain that can make training downstream components easier. 

 For our experiments, the language model was trained for $1,100,000$ steps over our dataset, using the default set of self-supervised language tasks used to train RoBERTa.
\section{Command Prediction}
\label{sec:command_prediction}

\begin{figure}[]
    \centering
    \includegraphics[width=\textwidth]{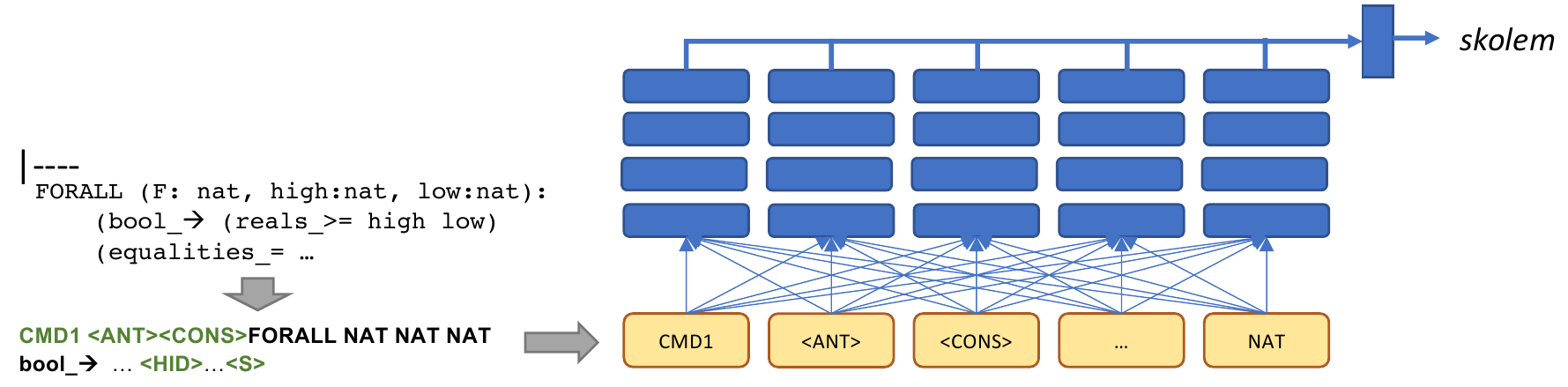}
    \caption{The process for featurizing a sequent and then using repeated self-attention to create representations capturing information for predicting the next command. }
    \label{fig:cmdpred_seq2seq}
\end{figure}

The goal of the command prediction task is to identify the next command an expert user may likely take given the current step in the proof.  
To make this prediction, we use the T5 sequence to sequence training framework \cite{raffel2019t5} implemented in Huggingface \cite{huggingface}.  Here, the RoBERTa encoding of the featurized proof state is used as input to output the command the expert user took given the proof state.
Figure \ref{fig:cmdpred_seq2seq} illustrates how a command prediction is made.  First the sequent is featurized into a token sequence, which are converted into classification-suitable representations via repeated applications of self-attention.  These are then integrated by the classifier, which writes out the next likely command.  As with most classifiers, the top-$N$ most confident hypotheses can be emitted, allowing for a window of predictions to be generated.

\section{Lemma Retrieval}
\label{sec:lemma_retrieval}
 \begin{figure}[h]
  \centering
  \includegraphics[width=0.75\textwidth]{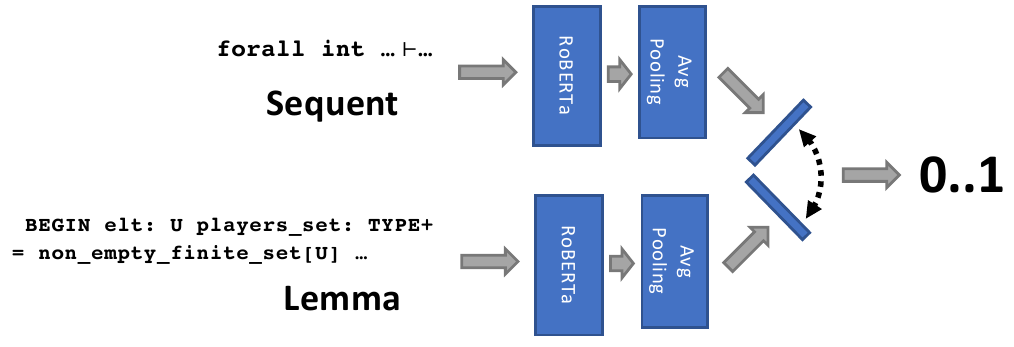}
  \caption{Siamese architecture used to determine whether a lemma is relevant to a given sequent.}
  \label{fig:sbert}
\end{figure}

As previously mentioned, we treat lemma retrieval as an information retrieval problem, where the proof state acts as the query against a library of available lemmas.
Given we have a large corpus of user-performed lemma imports in the PVSLib proof traces, we can use it to train a neural information retrieval approach \cite{intro_nir}, which learns a relevance ranking function from these examples.
CoProver takes this approach as the assumptions used to determine relevance between queries and natural language may not apply to matching proof states with lemmas.  
A supervised neural architecture can learn the best combination of features for assessing relevance of a lemma to a sequent.  
Figure \ref{fig:sbert} illustrates how the sequent and lemma are encoded and compared to assess relevance.  The architecture follows a Siamese Network \cite{siamese}, implemented in the SBERT \cite{reimers2019sbert} framework.  Given two inputs, Siamese networks encode them into a common representation which is scored typically by comparing the input representations with a tensor operation such as cosine similarity or dot-product.  Supervised training causes the architecture to learn feature combinations and weightings that are useful for computing the score.  When a fast operation such as cosine similarity is used to compute the score, comparisons can be extremely efficient as the bulk of the representations can be pre-computed.  This approach has been used to learn ranking functions for tasks with a large amount of data, such as using clickthrough data to rank queries and documents \cite{dssm_huang}.  
In CoProver, the token sequences for the logical formulas representing the lemma and sequent are passed to the RoBERTa encoder to construct token-level representations.  
These representations are averaged to give a single vector characterizing the content of the lemma or the sequent.  Finally, the relevance of the lemma and sequent vectors is scored using cosine similarity, with 0 indicating no relevance and 1 indicating maximal relevance.

\section{Experiments and Results}
\label{sec:experiments}

We detail the experimental setup and results for the Command Prediction and Lemma Retrieval tasks.

\subsection{Command Prediction}
From the full PVSLib library of proofs, we subsampled $20,000$ proof steps to create a tractable training set \footnote{Initial experiments with larger samples showed no difference in performance with a system trained with the smaller set}.  From these we randomly sampled $90\%$ of these pairs for the training data, and used the remaining $10\%$  as a held out test-set. 

For CoProver, we trained on this data for $10$ epochs on four NVIDIA GeForce RTX 3090 cards using distributed data parallel training (DDP) \cite{ddp} implemented using PyTorch Lightning \footnote{https://www.pytorchlightning.ai/}.  As each card in DDP contains its own model, we selected the one with the best validation set error.  
As baselines for comparison, we follow prior literature on tactic prediction \cite{gauthier2018tactictoe} and used classifiers trained over term-frequency inverse document frequency (TF-IDF) \cite{luhn57tf,jones72idf} weighted feature counts of symbols observed in the sequent.
TF-IDF incorporates frequency of occurrence of a term and its distinguishability against the backdrop of the entire collection.    
Here, the sequents are represented as a vector, with each dimension corresponding to a symbol in the vocabulary and weighted by TF-IDF.  In order to make a fair comparison, these were derived from the same featurized token stream used by CoProver.  To strengthen this type of approach, we experimented with multiple classifiers for the baseline: Linear support vector classifier (Linear SVC), support vector machines using a radial basis kernel (RBF) and one using a polynomial kernel (Poly), and a k-nearest neighbor classifier (k-NN).  We used the Scikit-Learn\footnote{https://scikit-learn.org/} implementations with default parameters unless specified.  For the k-nearest neighbor classifier, we used a distance weighted variant with $n=5$ following prior literature \cite{gauthier2018tactictoe}.

\begin{table}
    \centering
    \caption{Command prediction test accuracies by method, with and without command history information.}
    \begin{tabular}{c|c|c|c}
    \toprule
        \textbf{Method} & \textbf{Acc. cmdhist + sequent} & \textbf{Acc., sequent only} & \textbf{Acc., cmdhist only} \\
        \midrule
         Linear SVC &  $0.30 \pm 1.1 \times 10^{-2}$ & $0.20 \pm 9.1 \times 10^{-3} $ & $0.30 \pm 1.1 \times 10{-2}$ \\
         SVM (RBF) & $0.29 \pm 1.0\times 10^{-2}$ & $0.22 \pm 9.5 \times 10^{-3}$ & $0.30 \pm 1.0 \times 10^{-2}$\\
         SVM (Poly) & $0.20 \pm 8.9 \times 10^{-3}$ & $0.20 \pm 8.9 \times 10^{-3}$ & $0.22 \pm 1.0 \times 10^{-2}$\\
         k-NN & $0.28 \pm 1.0 \times 10^{-2}$ & $0.19 \pm 8.6 \times 10^{-3}$ & $0.27 \pm 9.6 \times 10^{-3}$ \\
         CoProver & $\mathbf{0.48 \pm 7.3 \times 10^{-3} }$ & $0.28 \pm 9.8 \times 10^{-3}$ &  $0.21 \pm 9.3 \times 10^{-3}$ \\
         \bottomrule
    \end{tabular}    
    \label{tab:my_label}
\end{table}

Table \ref{tab:my_label} shows the test command predication accuracy for each of the methods with the sequent and command history, and without the command history.  We find that the CoProver predictions are more accurate than the baselines when the full sequent and command histories are used.  Of note is the fact that most of the accuracies for the baseline methods is from the command history itself, whereas CoProver is able to integrate the sequent and command history together to achieve a significantly higher value than the next-best baseline, k-NN. Variances for each method were estimated using bootstrap resampling \cite{bootstrap} and significance was determined using a two-sample $t$-test with $\alpha = 0.001$.

\begin{table}
\centering
\caption{Command prediction accuracy for baseline methods using  features that incorporate more structural information (left to right).}
\begin{tabular}{c|c|c|c}
\toprule
\textbf{Method} & \textbf{n=$1$} & \textbf{n=$2$} & \textbf{n=$3$}\\
\midrule
Linear SVC &  $0.30$ & $0.37$ & $0.30$ \\
SVM (RBF) & $0.29$ & $0.32$ & $0.33$\\
SVM (Poly) & $0.20$ & $0.18$ & $0.19$ \\
k-NN & $0.28$ & $0.30$ & $0.32$ \\
\bottomrule
\end{tabular}

\label{tab:structural}
\end{table}

In order to determine the significance of structural information for this task, we experimented with TF-IDF sequent featurizations of increasing maximum $n$-gram degree.  Here, a $n$-gram featurization consists of all symbol sequences of length $n$.  At $n=1$, only single symbols are used and thus no structural information is encoded.  In contrast, $n=3$ uses all observed combinations of up to three symbols in sequence and contains considerably more of the sequent structure.  
Table \ref{tab:structural} shows the accuracies for each classification method by the maximum $n$-gram degree.  With the exception of the SVM using the polynomial kernel (SVM Poly), every method benefits from increasing structural information.  We suspect that model's poorer performance may be due to the greater number of hyperparameters given the polynomial kernel, which greatly increases the risk of overfitting on sparse data \cite{libsvmTutorial}.

\begin{figure}[h]
  \centering
  \includegraphics[width=0.8\textwidth]{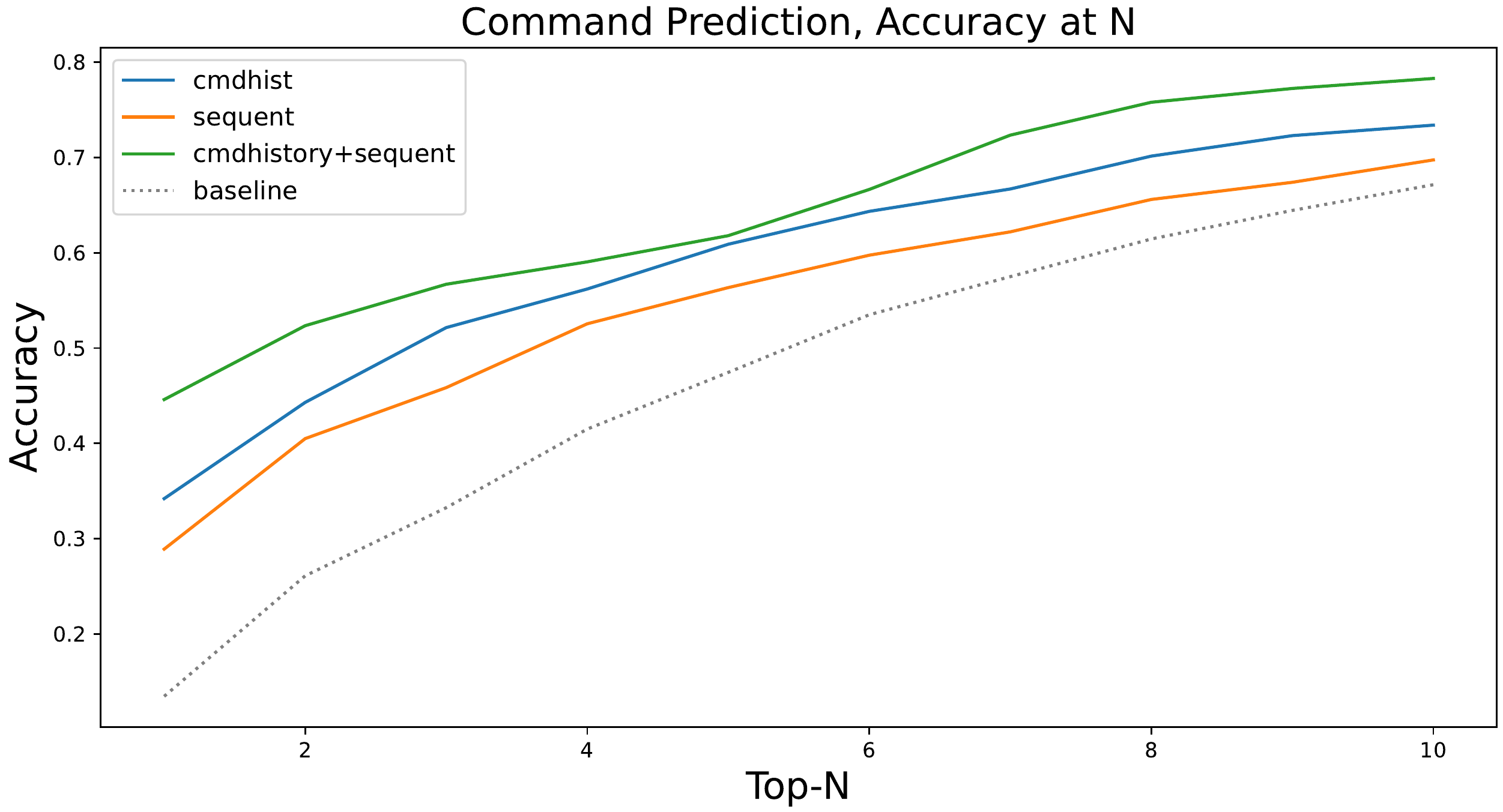}
  \caption{Command prediction test accuracies by method, with and without command history information. }
  \label{fig:cmdpred_topN}
\end{figure}

Recommender systems often present the top $N$-most relevant predictions for the user, instead of presenting just the top recommendation.
In many domains users can easily scan a set of candidates and choose the best one.  Alternately, in other cases the context provided by multiple candidates can provide useful information as well.
To assess how well CoProver can perform in this regime, we assess the top-$N$ test set accuracy, where matches are made if the correct prediction is within the top $N$ predictions.  Figure \ref{fig:cmdpred_topN} shows the accuracy of the CoProver system at different sizes of $N$ using just the command history (\texttt{cmdhist}), sequent (\texttt{sequent}) and both (\texttt{cmdhist+sequent}) for $N$ ranging from $1$ to $10$.  For comparison, we also include the baseline guess, where the top-$N$ most frequent commands in the training set are used as the candidate window.  We find that the combination of sequent and command history information gives consistently higher accuracy than using either alone, while all methods outperform the baseline.  Observe that the accurate command is present in the top-7 response over 70\% of the time.

\subsection{Lemma Retrieval}
For lemma retrieval, we winnowed the PVSLib proof steps to \term{lemma} commands that imported a lemma into the proof.  We restricted our analysis to explicit imports: While there are several other commands that also import lemmas, they do so as an implicit part of their operation and may have additional side effects.  From these $20,221$ steps, we randomly selected $12,132$ of them for training lemma retrieval models, with $8,089$ for testing.  These proofs reference libraries in both PVSLib and PVS Prelude, giving a total of $9,468$ possible candidate lemmas.

This model is trained on \verb!lemma! commands in the training set of proof sequences in the PVSLib.  For each \verb!lemma! invocation, we record the sequent at that point in the proof and the name of the referenced lemma.  PVSLib has $20,221$ \term{lemma} pairs in the proofs for PVSLib.  These consist of the sequent state when the import lemma command was issued by the user and the name of the lemma.  We randomly split them into $12,132$ train and $8,089$ test pairs.  This is against a combined library of PVS and PVSLib theories, with a total of $747$ theories and $9468$ available lemmas.

For training the Siamese network, the observed sequents and lemmas in the training invocations have a score of $1$.  An equal number of sequent and lemma pairs were sampled, with their score set to $0$, following the rationale that a randomly selected lemma would very likely not be relevant for progressing for a given sequent.  

We evaluated the resulting network on the test pairs.  We measured how well each method performed using mean reciprocal rank (MRR), an evaluation measure commonly used in information retrieval to assess how highly compared methods can rank relevant documents from a corpus \cite{manning2008introduction}.  MRR is computed from the rank position of the ground truth lemma in the relevance-ordered scores given the sequent.  For a given test pair consisting of a sequent $s$, selected lemma $l_{gt}$ and library of lemmas $L$ with $l_{gt} \in L$, we use the Siamese network to score how relevant lemma $l\in L$ is to the sequent, $f_{rel}(s, l)$.  We derive a rank ordering over all lemmas $L$, 

\begin{equation*}
    MRR = \frac{1}{N} \sum_i^{N} \frac{1}{r_i}
\end{equation*}

Where $r_i$ is the rank of the ground truth lemma for the $i^{th}$ lemma pair.  Thus higher MRR values indicate the scoring method does a better job of assigning higher relevance to ground truth documents.

\begin{center}
\begin{tabular}{c|c}
    \toprule
     \textbf{Method} &  \textbf{MRR} \\ 
     \midrule
     Baseline & $0.0015$ \\
     Count &  $0.0030$ \\
     TF-IDF & $0.043$ \\
     CoProver & $\mathbf{0.51}$ \\
     \bottomrule
\end{tabular}
\end{center}

We find that the CoProver approach to outperform the other methods for identifying relevant lemmas with their sequents, including the TF-IDF bag-of-words representation used by previous work.  A MRR of $0.51$ corresponds to an average mean rank of $1.98$, which corresponds to the relevant lemma appearing around position $2$ in a score-based rank ordering of all lemmas.  

%
\section{Deployments}
\label{sec:deployments}

CoProver is currently available as a web-application backed services for PVS.  We took the command prediction and lemma retrieval models and wrapped these up as web services in the Flask\footnote{\url{https://flask.palletsprojects.com/}} framework.  To request a command prediction or a relevant lemma, the calling component converts the current sequent into a JSON formatted message.  Hypotheses are returned as JSON objects, which are presented in PVS directly.

We are preparing this recommender-augmented version of PVS for a large user study to assess how useful these capabilities are to end users.  However, internal developers have found the capability to generally give good suggestions, when presenting the top $5$ commands.  

The code and the data used for this work are open-sourced and will be available at \url{https://github.com/coproof}.
%

\section{Related Work}
\label{sec:related_work}

In the last decade, there has been a significant amount of activity in applying machine learning to automated deduction.
The progress is summarized in an invited talk by Markus Rabe (Google Research) at the 2021 CADE conference~\cite{rabe2021towards}.   Prior work on using machine learning in theorem proving can be classified in terms of the predictive goal of machine learning as
\begin{enumerate}
\item 	Learning search heuristics (E-prover~\cite{Schulz:AICOM-2002}, SAT/SMT solvers~\cite{sat-handbook}): The systems ENIGMA~\cite{jakubuuv2017enigma}, MaLeCoP~\cite{urban2011malecop} and FEMaLeCoP~\cite{kaliszyk2015femalecop} augment a tableau-based prover LeanCoP~\cite{otten2003leancop} with a na\"{\i}ve Bayes classifier for clause selection.  Graph neural nets have been used to predict the clauses that are in the unsatisfiable core set of clauses from a clause set~\cite{selsam2019guiding} and to guide SMT solvers~\cite{balunovic2018learning}\@. 

\item Premise selection from a library of facts (DeepMath~\cite{irving2016deepmath}, CoqHammer, HOLyHammer, HOList~\cite{bansal2019holist}, Thor~\cite{https://doi.org/10.48550/arxiv.2205.10893}): Several proof assistants invoke hammers (theorem provers and SAT/SMT solvers) on each subgoal together with a set of background lemmas (the premises).  These hammers can fail if there are too many premises. Machine learning has been used to identify the most promising premises to pick from the background library~\cite{kuhlwein2013mash}. As with tactic selection below, a range of learning techniques have been employed for premise selection, including sequence, tree, and graph representations~\cite{wang2017premise}.
  
\item 	Step or tactic selection (GPT-f, Holophrasm~\cite{whalen2016holophrasm}, CoqGym~\cite{yang2019learning}, HOL4RL~\cite{wu2021tacticzero}, HOList~\cite{loos2017deep}, GamePad~\cite{huang2018gamepad}, Tactic-Toe~\cite{gauthier2018tactictoe}, proof synthesis~\cite{first2020tactok,https://doi.org/10.48550/arxiv.2210.12283,https://doi.org/10.48550/arxiv.2205.11491}):  Interactive proof assistants build proof trees by applying tactics to goals to generate zero or more subgoals.   The SEPIA system~\cite{gransden2015sepia} predicts tactics for the Coq proof assistant based purely on analyzing proofs.  GPT-f uses the GPT-3 transformer model to train on Goal/Proof pairs from the Metamath corpus (augmented with synthesized proofs) to predict the proof given the goal.  This is used to construct a proof tree by applying the proof steps suggested by the model to the open subgoals in the tree.  The system was able to find shorter proofs for 23 theorems in the Metamath corpus.  CoqGym~\cite{yang2019learning} uses a much larger training corpus spanning 71,000 proofs from various Coq libraries.  TacticToe~\cite{gauthier2021tactictoe} is trained on proofs from HOL4 libraries and combines tactic prediction using k-nearest neighbors with A* search, and in some cases yields better (more perspicuous and maintainable) proofs than the alternatives using Hammers.  IsarStep~\cite{li2020isarstep} uses a transformer encoding to identify intermediate formulas in a declarative Isar proof.  
ML4PG~\cite{komendantskaya2012machine} extracts useful statistical patterns from higher-order logic proofs using unsupervised learning techniques like K-Means clustering.   
\end{enumerate}

Learning to predict the next command from user activity traces has been examined in works such as TacticToe \cite{gauthier2018tactictoe}.  However, the focus of this work was on the ability of the system to fully automate the proof, and performance was measured in terms of the lift in number of proofs that can be completed without human assistance.  To our knowledge, prior work has not examined how well these systems can predict the next step a user would take to make progress.

In addition, previous work has also featurized sequents using a ``bag-of-words'' approach, where the proof state is characterized as a histogram of the tokens observed in the sequent.  While pairs of histograms of pairs of adjacent tokens can retain some sequentiality, this approach can ablate longer-range structural information that may be valuable for tasks such as prediction.  In contrast, CoProver leverages neural language modeling technology that can learn sequential combinations of tokens over longer ranges.  

  Treating lemma retrieval as an information retrieval problem has been done in prior work \cite{blanchette2016factselector,gauthier2018tactictoe}.  As with most IR methods, these works use a weighted histogram of terms observed in the sequent (query) and lemma (document) to characterize them.  While this ``bag-of-words'' approach removes sequentiality and thus structural information, it turns out the overlap in vocabulary use between a query and candidate document can act as a good approximator for search relevance.  Terms are relevance weighted using term-frequency inverse document frequency (TF-IDF) \cite{luhn57tf,jones72idf}, which incorporates frequency of occurrence of a term and its distinguishability against the backdrop of the entire collection.  The similarity comparison is usually performed using a fast computational method, such as strict dot products or cosine similarity, so a query can be tractably performed against a large collection of candidate documents.

However, bag-of-words modeling and the TF-IDF weighting scheme are assumptions targeting how relevance appears for natural language queries and documents.  Logical formula observed in lemmas and sequents may not exhibit the same behavior, particularly for determining if a lemma is relevant to moving proof progress in a sequent.  Indeed, neural information retrieval has focused on using supervised queries and document pairs to learn relevance functions that may not be captured by assumptions taken in standard IR modeling.
\section{Conclusions}
\label{sec:conclusions}

In this work, we have demonstrated how a simple featurization of proof state can be used to perform two recommendation tasks, predicting next commands and retrieving relevant lemmas.  
For command prediction, CoProver's approach has been shown to outperform prior methods, giving significantly higher accuracies.  In the context of recommendation systems, showing the top $3-5$ commands is a reasonable amount, with these windows capturing $50\%-70\%$ of the original prediction correctly on the validation set.  As with systems trained on user interaction traces, there will are often many cases where the system can learn a solution that the original user did not consider.  As one internal user commented, the application of an automated hammer (the \term{grind} command) in a convergence proof was unexpected, but lead to completion of the proof.  Similarly, using a neural learning approach with CoProver's featurization can give significantly better performance on lemma retrieval, in comparison with retrieval using IR-derived baselines.  

In spite of these results, we note that the neural learning mechanism are not necessarily learning deep reasoning structures, and may more likely be learning complex structural cues.  Indeed, an analysis of large language models found them to be impressive memorization machines that are incapable of performing arithmetic \cite{LLMs_fewshot_learners}.  A cursory examination of the attention heads in the command prediction task revealed the model's attention weightings did not consistently align with experienced users' intuitions about what should govern the direction of the proof.  

Future directions of proof command recommendation include identification of arguments used for these commands.  These primarily consist of the formula to use, but in some cases more complex arguments are needed.  Of particular interest is pairing this capability with explanation mechanisms.  Perhaps the simplest explanation capability is to run the top-N commands in the background and displaying the results provides a look ahead capability that allows users to see the envelope of outcomes.  When paired with heuristics that measure proof completion, this may be beneficial for developing an intuitive understanding.

To the best of our knowledge, we are the first work to assess and evaluate lemma retrieval as an information retrieval problem.  Previous work on lemma retrieval focused on whether or not a selected lemma can progress a proof towards completion in an automated solver.  While this does cover situations where multiple lemmas may be relevant to automatic proof completion, for the type of problems addressed by interactive theorem provers, automatic completion may not be feasible.  However, importing a lemma can help progress the proof and thus can be considered relevant, similar to how retrieving the right document can help a querying user perform a task.

To that end, we have demonstrated how a neurally trained architecture can determine which lemma a human user would have selected for making progress on a proof given the sequent.  This approach can provide a better relevance ranking for lemmas, as opposed to the representation and scoring methods discussed in previous work.  

Note that from our data setup, only the lemmas selected by the user are considered to be relevant for the given sequent.  This disregards the possibility that another lemma may be just as useful for the proof as well.  However, this is a well known issue in natural language information retrieval corpora, and measures such as MRR are used to compare system performance as opposed to acting as a standalone performance measure.

Possible future work in this area can focus on analyzing relevant structural elements that trigger a match between a sequent and a lemma.  While the final comparison is performed using a cosine similarity computation, the nature of the highest scoring feature matches can be hard to discern.  In particular, it is possible that the formula for the lemma and sequent may not have much apparent overlap, but relevant token sequences may map to the same feature.

For future work, we are examining the application of CoProver's Transformer based sequent representation towards tasks such as nominating witnesses, proof repair, and developing a measure for proof progress.
\subsubsection{Acknowledgements} This material is based upon work supported by the Defense Advanced Research Projects Agency (DARPA) under Contract No.\ HR00112290064 and by the National Institute of Aeronautics.  Any opinions, findings, and conclusions or recommendations expressed in this material are those of the author(s) and do not necessarily reflect the views of the United States Government or DARPA.    

\bibliographystyle{splncs04}
\bibliography{main}
\end{document}